\documentstyle[aps,pre,epsfig,preprint]{revtex}

\begin{document}
\pagestyle{myheadings} \markboth{Shvetsov/Helbing: Macroscopic Dynamics of
  Multi-Lane Traffic} {Shvetsov/Helbing: Macroscopic Dynamics of
  Multi-Lane Traffic} \title{Macroscopic Dynamics of Multi-Lane Traffic}
\author{Vladimir Shvetsov and Dirk Helbing} \address{II. Institute of
  Theoretical Physics, University of Stuttgart,\\ Pfaffenwaldring 57/III,
  70550 Stuttgart, Germany,\\ 
  {\tt http://www.theo2.physik.uni-stuttgart.de/helbing.html}}
\maketitle \draft
\begin{abstract}
  We present a macroscopic model of mixed multi-lane freeway traffic that
  can be easily calibrated to empirical traffic data, as is shown for
  Dutch highway data. The model is derived from a gas-kinetic level of
  description, including effects of vehicular space requirements and
  velocity correlations between successive vehicles. We also give a
  derivation of the lane-changing rates. The resulting dynamic velocity
  equations contain non-local and anisotropic interaction terms which
  allow a robust and efficient numerical simulation of multi-lane traffic.
  As demonstrated by various examples, this facilitates the investigation
  of synchronization patterns among lanes and effects of on-ramps,
  off-ramps,
  lane closures, or accidents.\\
\end{abstract}

\section{Introduction}

Simulating traffic flow is not only of practical importance for developing
traffic optimization measures \cite{Bovy}. It is also interesting because
of the various self-organization phenomena resulting from non-linear
vehicle interactions \cite{TGF95,TGF97,Buch}. This includes the formation
of traffic jams or stop-and-go waves \cite{KK-93,KK-94,Hel-95},
and of synchronized congested traffic \cite{KK-96,KK-97,HT-98}.  Many of
these phenomena can be simulated by one-dimensional models like cellular
automata \cite{NS,Barlovich} and macroscopic traffic models
\cite{KK-93,THH-98}.  It is, however, questionable if traffic dynamics at
on- and off-ramps or intersections can be correctly described without
consideration of lane changes
\cite{GaHeWe62,MuPi71,MuHs71,Ror76,MaNa81,MiBe84}.  The same holds close
to changes in the number of lanes, accidents, or construction sites. It
is, therefore, desireable to have multi-lane models that explicitly take
into account overtaking maneuvers and lane changes. In contrast to
existing cellular automata models for multi-lane traffic
\cite{CAML1,CAML2,CAML3,HH-98}, we will focus on macroscopic models, here,
since they allow analytical investigations and a simple calibration
to empirical data.
\par
Apart from lane-changing maneuvers, traffic dynamics is considerably
influenced by the composition of traffic into various types of vehicles
with different desired velocities and acceleration capabilities.  This can
even cause new kinds of phase transitions in mixed traffic, e.g. to a
coherent, solid-like state of motion \cite{HH-98}.  Thus, it is also
favourable to have a macroscopic model that allows to distinguish several
vehicle types.
\par
We will obtain such a model from a generalized version of a gas-kinetic
traffic model, from which we managed to derive a one-dimensional model
that is consistent with all presently known properties of traffic flow
\cite{THH-98}, including synchronized congested flow \cite{HT-98}.
Although lane-changes and several vehicle types make the model quite
complex, it is still possible to evaluate the Boltzmann-like interaction
terms. Moreover, whereas previous approaches have neglected
correlations between successive vehicles, we will also show how these can
be taken into account.
\par
Our calculations take care of the fact that vehicles do not interact
locally, but with the next vehicle in front, so that the interaction point
is advanced by about the safe vehicle distance. Nevertheless, we were able
to evaluate the interaction integral without the gradient expansion used
in previous publications \cite{Physica,High}. This resulted in
characteristic non-local and anisotropic interation terms, which have very
favourable numerical properties \cite{numerics}. 
Like diffusion or viscosity terms, they
cause a smoothing of shock fronts. However, they do not change the
character of the set of partial differential equations from a hyperbolic
to a parabolic one. Finally, the non-local terms allow to write the
macroscopic traffic equations in flux form with source/sink terms, so that
numerical standard procedures can be used for their robust and efficient
numerical integration \cite{numerics}. 
As a consequence, it is now possible to solve the
multi-lane traffic equations without elimination of the dynamic velocity
equations, i.e. without averaging over self-organized velocity variations
like stop-and-go traffic. This had to be done in previous approaches
because of numerical stability \cite{ML3}, with the consequence that the
investigation of unstable multi-lane traffic was not possible.  Now, we
are able to study how the vehicle dynamics on one lane influences the
others.
\par 
Our paper is organized as follows: In Section~II we will introduce the
kinetic model, which allows the derivation of the macroscopic
multi-lane model of heterogeneous traffic presented in Section~III.
The decisive steps of this derivation are given in the Appendix.
Section~IV will discuss the calibration of the model to real traffic
data for the case of one effective vehicle type, and it will display various
simulation results for difficult test scenarios like lane closures.
A comparison with results of an effective single-lane model will be the
topic of Section V, while Section~VI summarizes the results of this paper.

\section{The Kinetic Multi-Lane Model}

Prigogine and coauthors \cite{prig} were the first who
proposed a kinetic equation for the
phase-space density of vehicles on a highway. Later, 
Paveri-Fontana \cite{paveri}
made important improvements, formulating an equation for the
``extended'' phase-space density $g(x,v,v_0,t)$, where $x$ is the location,
$v$ the {\em
  actual} velocity, and $v_0$ the {\em desired} velocity of a vehicle
at time $t$.
Nagatani and Helbing have suggested extensions to multi-lane traffic
\cite{ML3,Naga}. However, the
validity of all theses equations was restricted to the regime of small
densities.

Here, we present the general kinetic model for multi-lane traffic composed
of different types of spatially extended vehicles, and derive the
corresponding
macroscopic model from it. The main features of the model are
\begin{itemize}
\item the correct description of dense traffic by taking into account 
the finite space requirements of vehicles,
\item the ``non-locality of the interactions'', which means that drivers look
  ahead and adopt their behavior to the traffic situation at some position
  in front of them,
\item the replacement of the ``vehicular chaos'' assumption used by 
  most previous
  kinetic traffic models \cite{Physica,High,prig,paveri,Wagner}
  (with a few exceptions \cite{Nelson,Klar}) by
  much less restrictive assumptions, which
  account for possible velocity correlations of interacting vehicles.
\end{itemize} 
For simplicity, we assume that all vehicles of the same type 
$a \in \{1,\dots,A\}$ have the same
desired velocity $V_{0i}^a$ that may depend on the lane
$i\in \{1,\dots,N\}$. Then, we can represent the ``extended''
phase-space density $g_i(x,v,v_0,t)$
of all vehicles in each lane through a sum of phase-space
densities $f_i^a(x,v,t)$ of vehicles of different types $a$:
\begin{equation}
g_i(x,v,v_0,t) = \sum_a f_i^a(x,v,t) \delta(v_0-V_{0i}^a) \, .
\end{equation}

The phase-space densities obey the following continuity equation 
\cite{Buch,ML3}:
\begin{equation}
  \frac{\partial f_i^a}{\partial t} +
  \frac{\partial}{\partial x} (f_i^a v) +
  \frac{\partial}{\partial v} \left( f_i^a \frac{V_{0i}^a-v}{\tau_i^a} \right) =
  \left(\frac{\partial f_i^a}{\partial t}\right)_{\rm int} +
  \left(\frac{\partial f_i^a}{\partial t}\right)_{\rm lc} \, .
\end{equation}
The terms on the left-hand side represent the continuous change of the
phase-space density due to the movement of a point in phase space. The
third term represents the tendency of drivers to accelerate to their
desired velocity $V_{0i}^a$
with a relaxation time $\tau_i^a$. The terms on the
right-hand side represent (quasi-)discontinuous changes of the phase-space
densities due to lane-changing maneuvers and braking interactions between 
vehicles. 

We consider three different types of lane-changing behaviour:
\begin{itemize}
\item lane-changing maneuvers due to interactions between vehicles, which
  correspond to overtaking maneuvers or lane changes in order to avoid
  intensive interactions with slower vehicles in a lane,
\item ``spontaneous'' lane-changing maneuvers, which reflect effects of
  traffic regulations and the internal tendency of drivers to use a
  particular lane,
\item ``mandatory'' lane changes due to lane mergings, on- or
  off-ramps, accidents etc.
\end{itemize}
Accordingly, we have:
\begin{equation}
  \left(\frac{\partial f_i^a}{\partial t}\right)_{\rm lc} =
  \left(\frac{\partial f_i^a}{\partial t}\right)_{\rm lc}^{\rm int} +
  \left(\frac{\partial f_i^a}{\partial t}\right)_{\rm lc}^{\rm spont} +
  \left(\frac{\partial f_i^a}{\partial t}\right)_{\rm lc}^{\rm mand} \, .
\end{equation}

Following Prigogine, we speak of ``interaction'', when a faster vehicle
approaches a slower one, which forces the faster vehicle to change lane 
or brake. For simplicity, we will assume that
vehicles do not change their velocities during lane-changing
maneuvers and that braking vehicles decelerate exactly to the velocity of
the vehicle in front.

The key quantities, which determine the RHS of the kinetic
equations, are the ``interaction rates'', or the effective number of
interactions between vehicles per unit of time. Let us
denote by ${\mathcal I}_i^{ab}(x,v,t)$ the interaction rate of 
vehicles of type $a$ and velocity $v$ at place $x$ 
in lane $i$ with slower vehicles of type $b$ and velocity $w<v$ in front; 
by ${\mathcal J}_i^{ab}(x,v,t)$ the number of interactions
in lane $i$ between vehicles of type $a$ with velocity $w>v$ 
and vehicles of type $b$ in front with velocity $v$. 
The interaction rates ${\mathcal I}_i^{ab}(x,v,t)$ with
$b\in\{1,\dots,A\}$ contribute to the decrease of the 
phase-space density $f_i^a(x,v,t)$ due to lane changing or braking, whereas
${\mathcal J}_i^{ab}(x,v,t)$ contribute to the increase of the 
phase-space density due to braking of vehicles of type $a$ to 
velocity $v$. They are given by the following formulas:
\begin{eqnarray}
  {\mathcal I}_i^{ab} &=& \chi_i\int\limits_{v>w} dw \; (v-w)
  f_i^{ab}(x,v,x+s_i^a,w,t) \, , \label{eq_intrate1}\\
  {\mathcal J}_i^{ab} &=& \chi_i\int\limits_{w>v} dw \; (w-v)
  f_i^{ab}(x,w,x+s_i^a,v,t) \, .\label{eq_intrate2}
\end{eqnarray}
That is, the interaction rates are proportional 
\begin{itemize}
\item to the ``effective cross section'' $\chi_i=\chi_i(\rho_i)$, 
reflecting the increased number of interactions in dense traffic due
to the vehicular space requirements (see below), 
\item to the
relative velocities $|v-w|$ of interacting vehicles, and 
\item to the
pair distribution function $f_i^{ab}(x,v,x+s_i^a,w,t)$, describing the
phase space density of having in lane $i$ two vehicles of types $a$ and $b$
at places $x$ and ${x_i^a}'=(x+s_i^a)$ 
with velocities $v$ and $w$, respectively. Note that
the classical ``vehicular chaos'' assumption would correspond to the
factorization $f_i^{ab}(x,v,x+s_i^a,w,t) = f_i^a(x,v,t)f_i^b(x+s_i^a,w,t)$
of the pair distribution function into the phase space densities
of single vehicles.
\end{itemize}
With $s_i^a>0$ we assume an anticipative driver behavior, i.e. a reaction
to the traffic situation at the advanced ``interaction point''
${x_i^a}' = (x+s_i^a)$. A reasonable formula is $s_i^a 
= \gamma_i^a (1/\rho_i^{\rm max} + T_i^a V_i)$,
where $\rho_i^{\rm max}$ is the maximum vehicle density in lane $i$, $T_i^a$ is
the safe time headway of vehicles of type $a$, 
and $T_i^a V_i$ is the safety distance at an average
velocity of $V_i$ in lane $i$. For $\gamma_i^a = 1$, the vehicles react to the
traffic situation at the safe vehicle distance, for $\gamma_i^a >1$, they
look further ahead. The anticipation factor $\gamma_i^a$ is typically
between 1.0 and 3.0 \cite{THH-98}.

Now, let us denote by $p_i^a$ the probability that a vehicle of type $a$ in
lane $i$ can change lane without any delay. Evidently, this probability
equals the sum of probabilities of changing to the right lane
$(i-1)$ or to the left lane $(i+1)$: 
$p_i^a = p_{i,i-1}^a + p_{i,i+1}^a$. We assume these probabilities
to be functions of the macroscopic variables such as vehicle densities.

In terms of the interaction rates, 
we can now specify the rate of braking interactions by
\begin{equation}
  \left(\frac{\partial f_i^a}{\partial t}\right)_{\rm int} =
  (1-p_i^a)\sum_b \Big[ 
 {\mathcal J}_i^{ab}(x,v,t) - {\mathcal I}_i^{ab}(x,v,t) \Big] \;,
\end{equation}
and the lane-changing term due to interactions by
\begin{eqnarray}
  \left(\frac{\partial f_i^a}{\partial t}\right)_{\rm lc}^{\rm int} &=& \sum_b
   \Big[ p_{i-1,i}^a {\mathcal I}_{i-1}^{ab}(x,v,t) -
    p_{i,i-1}^a {\mathcal I}_i^{ab}(x,v,t) \nonumber \\ & & \quad +
    p_{i+1,i}^a {\mathcal I}_{i+1}^{ab}(x,v,t) -
    p_{i,i+1}^a {\mathcal I}_i^{ab}(x,v,t) \Big] \, .
\end{eqnarray}

The rates of spontaneous lane-changes are simply proportional to the
phase-space densities of vehicles. The proportionality factors are
determined by the transition rates $1/T_{i-1,i}^a$
of this process, where $T_{i-1,i}^a$ are the characteristic times
between lane changes. We will assume that they are functions of the
macroscopic variables like the densities. 
Thus, the spontaneous lane-changing terms have the form:
\begin{eqnarray}
  \left(\frac{\partial f_i^a}{\partial t}\right)_{\rm lc}^{\rm spont} &=&
  \frac{f_{i-1}^a(x,v,t)}{T_{i-1,i}^a}  -
  \frac{f_i^a(x,v,t)}{T_{i,i-1}^a} \nonumber \\ &+& 
  \frac{f_{i+1}^a(x,v,t)}{T_{i+1,i}^a}  -
  \frac{f_i^a(x,v,t)}{T_{i,i+1}^a} \, .
\end{eqnarray}
An appropriate specification of the transition rates is the following:
\begin{equation}
  \frac{1}{T_{i,j}^a} = g_{i,j}^a\left(\frac{\rho_i}{\rho_i^{\rm max}}
  \right)^{\beta_1}
  \left(1-\frac{\rho_j}{\rho_j^{\rm max}}\right)^{\beta_2}. \label{eq-sparmann}
\end{equation}
This form is rather simple and, at the same time, it is in accordance with
empirical data \cite{Sparmann}. The results displayed in this paper are
for $\beta_1=0$ and $\beta_2=8$. 

The form of the mandatory lane-changing term depends on many different factors
including visibility or the location of traffic signs indicating the
end of a lane. An explicit example will be given in Section IV.

\section{The Macroscopic Multi-Lane Model} \label{sec-macro}

One advantage of the kinetic equation is, that it allows the
systematic derivation of equations for {\em macroscopic} variables. The
macroscopic variables of interest are the densities $\rho_i^a(x,t)$,
average velocities $V_i^a(x,t)$, and velocity variances $\theta_i^a(x,t)$ of
vehicles of type $a$ in lane $i$ at place $x$ and time $t$. 
They can be defined as moments of the phase-space densities:
\begin{eqnarray}
  \rho_i^a(x,t)   &=& \int dv \; f_i^a(x,v,t)\, , \\
  V_i^a(x,t)      &=& {\rho_i^a(x,t)}^{-1} \int dv \; v f_i^a(x,v,t)\, , \\
  \theta_i^a(x,t) &=& {\rho_i^a(x,t)}^{-1} \int dv \; (v-V_i^a)^2 f_i^a(x,v,t)
  \, . 
\end{eqnarray}

One can obtain the macroscopic equations from the kinetic equation, using
an iterative procedure, called Chapman-Enskog expansion 
\cite{Buch,drygran}. The first
step of this procedure gives the so-called Euler-like traffic equations,
which are in good agreement with empirical findings \cite{HT-98,THH-98}.
To obtain these, one assumes the velocity distribution in every point $x$
to be in a local equilibrium. Although the exact equilibrium solution of
the kinetic traffic equations is not known, the Gaussian distribution of
velocities is a good approximation \cite{fund}. 
Next, recall that we need to specify
the form of the pair distribution function of interacting vehicles. Previous
models \cite{Physica,High,prig,paveri,Wagner}
assumed the pair distribution function to be a product of 
Gaussian one-vehicle distribution functions, which corresponds to the 
``vehicular chaos'' assumption that
interacting vehicles are uncorrelated. Here, we only
presuppose that the pair distribution function can be approximated by 
a general bivariate Gaussian distribution function:
\begin{equation}
  f_i^{ab}(x,v,x+s_i^a,w,t) =
  \rho_i^a(x,t)\rho_i^b(x+s_i^a,t)\frac{\sqrt{\mbox{det}\mathcal{B}}}{2\pi}
  e^{-\frac{1}{2} {\mathcal B}(v,w)}\, , \label{eq-bivar1}
\end{equation}
where ${\mathcal B}(v,w)$ is a general positive definite quadratic
form and $\mbox{det}\mathcal{B}$ the determinant of the corresponding
symmetrical matrix. Thus, we take into account possible
correlations between the velocities of interacting vehicles.
One can express the coefficients of ${\mathcal B}(v,w)$ in terms of
the moments of the distribution, namely the variances $\theta_i^a$ and
the correlation coefficient $k_i^{ab}$:
\begin{equation}
{\mathcal B}(v,w) = \frac{1}{1-(k_i^{ab})^2}
\left(
  \frac{(v-V_i^a)^2}{\theta_i^a}
  -2 k_i^{ab}\frac{(v-V_i^a)(w-{V_i^b}')}{\sqrt{\theta_i^a{\theta_i^b}'}}
  +\frac{(w-{V_i^b}')^2}{{\theta_i^b}'} \right) \, . 
\label{eq-bivar2}
\end{equation}
The associated determinant equals 
\begin{equation}
  \mbox{det}{\mathcal B} = \left\{ \theta_i^a{\theta_i^b}'[1- (k_i^{ab})^2]
  \right\}^{-1} \, .
\end{equation}
A prime indicates that the respective quantity is
evaluated at the advanced interaction point ${x_i^a}'
=(x+s_i^a)$ rather than the
actual position $x$.

In order to evaluate the macroscopic equations, one needs to multiply the
kinetic equation by $v^k$, integrate over the velocity $v$,
and close the system of
equations by a suitable approximation for the moments
of higher order \cite{Buch,High,drygran}. We will derive the Euler-like
equations for the vehicle densities 
and average velocities, and close the system by
approximations for the variances and correlation coefficients.

The details of the calculation are given in the Appendix, here we present only
the final results.

The density equations are 
\begin{eqnarray}
  & & \hspace*{-6mm} \frac{\partial}{\partial t} \rho_i^a + 
  \frac{\partial}{\partial x} (\rho_i^a V_i^a) \nonumber \\ &=&
  \sum_{j\in\{i-1,i+1\}}\sum_{b=1}^A
  \left[
    p_{j,i}^a A_j^{ab}(\delta V_j^{ab}) - 
    p_{i,j}^a A_i^{ab}(\delta V_i^{ab})
  \right] \nonumber \\
  & & + \sum_{j\in\{i-1,i+1\}} \left(
    \frac{\rho_j^a}{T_{j,i}^a}  - \frac{\rho_i^a}{T_{i,j}^a} 
  \right) \, , \label{eq_den} 
\end{eqnarray}
and the equations for the traffic flows read:
\begin{eqnarray}
  & & \hspace*{-6mm} \frac{\partial}{\partial t}(\rho_i^a V_i^a) + 
  \frac{\partial}{\partial x}[\rho_i^a ({V_i^a}^2+\theta_i^a)] \nonumber \\
  &=&
  \frac{\rho_i^a}{\tau_i^a}(V_{0i}^a-V_i^a)
  - (1-p_i^a) \sum_{b=1}^A B_i^{ab}(\delta V_i^{ab}) \nonumber \\
  & & + \sum_{j\in\{i-1,i+1\}}\sum_{b=1}^A
  \Big[
    p_{j,i}^a C_j^{ab}(\delta V_j^{ab}) - 
    p_{i,j}^a C_i^{ab}(\delta V_i^{ab})
  \Big] \nonumber \\
  & & + \sum_{j\in\{i-1,i+1\}} \left(
    \frac{\rho_j^a V_j^a}{T_{j,i}^a}  - \frac{\rho_i^a V_i^a}{T_{i,j}^a} 
  \right). \label{eq_flux} 
\end{eqnarray}
One can obtain the corresponding velocity equations by inserting
(\ref{eq_den}) into~(\ref{eq_flux}). However, we will use the flux
equations~(\ref{eq_flux}) instead, because they are more suitable
for numerical integration methods.

The functions $A_i^{ab}$, $B_i^{ab}$, and $C_i^{ab}$ in the 
above equations are denoted as {\em
Boltzmann factors}, since they originate from the Boltzmann-like
interaction integrals~(\ref{eq_intrate1}) and (\ref{eq_intrate2}). Thus they
describe the influence of interactions on traffic dynamics. The Boltzmann
factors $A_i^{ab}$ determine the
lane-changing flows due to interactions in the density equations, $B_i^{ab}$ 
the braking term, and $C_i^{ab}$ the lane-changing terms due to interactions
in the flow equations. The exact
form of these terms is as follows (for brevity we omit indices of
lanes and vehicle types):
\begin{eqnarray}
  A(\delta V) &=& \chi(\rho)\rho\rho'\sqrt{S} 
  \left[
    {N}(\delta V) + \delta V {E}(\delta V) 
  \right] \, , \label{eq-boltz1}\\
  B(\delta V) &=& \chi(\rho)\rho\rho'S 
  \left[
    \delta V {N}(\delta V) + (1+\delta V^2) {E}(\delta V) 
  \right] \, , \label{eq-boltz2}\\
  C(\delta V) &=& \chi(\rho)\rho\rho'S \left[
    \frac{V}{\sqrt{S}} {N}(\delta V) +
    \left(
      \frac{\theta-k\sqrt{\theta\theta'}}{S}+\frac{V}{\sqrt{S}} \, \delta V
    \right)
    {E}(\delta V) \right] \, . \label{eq-boltz3} 
\end{eqnarray}
Here, ${N}(z)$ is the standard Gaussian distribution, and ${E}(z)$ 
denotes the error function:
\begin{eqnarray}
  {N}(z) &=& \frac{e^{-\frac{z^2}{2}}}{\sqrt{2\pi}} \,, \label{eq-norm} \\
  {E}(z) &=& \frac{1}{\sqrt{2\pi}} \int\limits_{-\infty}^z 
 dy \; e^{-\frac{y^2}{2}} \, . \label{eq-error}
\end{eqnarray}
The values $\delta V_i^{ab}$ are dimensionless velocity differences between
interacting vehicles, which are defined by
\begin{equation}
  \delta V_i^{ab} = \frac{V_i^a-{V_i^b}'}{\sqrt{S_i^{ab}}} \, , 
\label{eq-veldiff}
\end{equation}
where
\begin{equation}
  S_i^{ab} = \theta_i^a - 2 k_i^{ab} \sqrt{\theta_i^a{\theta_i^b}'} +
  {\theta_i^b}'. \label{eq-effvar}
\end{equation}
The Boltzmann factors are negligible for negative velocity differences,
and they grow rapidly with increasing positive differences. The
reason for this is rather intuitive: Faster vehicles in front do not influence
vehicles at the given place $x$, while slower vehicles force
them to brake or change lane. 

\begin{figure}
  \begin{center}
    \includegraphics[width=0.5\textwidth]{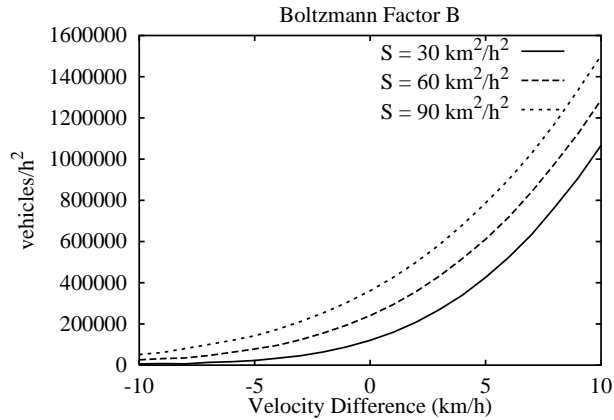}
  \end{center}
  \caption[]{ \label{fig-bfactor} Boltzmann factor $B$ as a function of 
the absolute velocity difference for different values of the variance
factor $S$. }
\end{figure}

The dimensionless values~(\ref{eq-veldiff}) have the meaning of 
``effective'' velocity
differences. According to their definition, they increase with 
\begin{itemize}
\item the increase
of the absolute velocity difference, 
\item the decrease of the variance, and 
\item the increase of the correlation coefficient.
\end{itemize}
The last two properties require some explanation. The increase of the
variance as well as the decrease of the correlation 
coefficient both lead to an
increase of the factor $S$. This results in the decrease of the effective
velocity difference and, consequently, of the dimensionless 
parts of the Boltzmann terms (standing in square brackets). 
Nevertheless, the Boltzmann factors themselves increase (see
Fig.~\ref{fig-bfactor}, which shows the dependence of 
Boltzmann factor $B$ on the absolute velocity difference for different
values of $S$). This effect has a clear interpretation: The 
dimensionless parts of the Boltzmann factors describe the influence of the
{\em difference in the average velocities} at location $x$ and the advanced
interaction point $x'$ on the number of interactions. 
An increase of $S$ reduces the effect of this difference in comparison 
with the high {\em variation of
individual velocities} in the vehicle flow. Thus, the increase of
$S$ diminishes the value of the effective velocity difference $\delta V$, 
although it enlarges the interaction rates. One could also say:
An increase of $S$ causes higher interaction rates, but also a wider
transition region between the limiting cases $(V_i^a-{V_i^b}') \ll 0$ and
$(V_i^a -{V_i^b}') \gg 0$.

For the variances $\theta_i^a$, we use constitutive relations of the form
\begin{equation}
  \theta_i^a = \alpha_i^a(\rho_i) {V_i^a}^2 \, ,
\end{equation}
according to which the variance of vehicle velocities is a certain
proportion $\alpha_i^a$ of the squared average velocity, which depends
on the total vehicle density
\begin{equation}
 \rho_i = \sum_{a=1}^A \rho_i^a
\end{equation}
in the respective lane $i$. This
is well justified by empirical findings. The appropriate expression of
the functions $\alpha_i^a(\rho_i)$ is given below.

The correlation coefficients could be approximated 
as functions of the densities
at the points $x$ and 
${x_i^a}'=(x+s_i^a)$, as well as the distance $s_i^a$: $k_i^{ab} =
k_i^{ab}(\rho_i^a, {\rho_i^b}', s_i^a)$. However, an empirical determination
of this function is very difficult, as it requires a
thorough analysis of a huge amount of single-vehicle data. 
Therefore, we will apply the common approximation 
$k_i^{ab}\approx 0$ for the time being. 

Another important function to be estimated is the ``effective cross section''
$\chi_i=\chi_i(\rho_i)$. This value reflects the increase of the
effective number of interactions in dense traffic. In a previous
publication on the single-lane variant of the above 
model~\cite{THH-98}, it was shown that the
following expression for the effective cross section is consistent with
the limiting cases at high and low vehicle densities and 
well justified by the resulting properties of the model:
\begin{equation}
  [1-p(\rho)]\chi(\rho) = \frac{V_0 T^2}{\tau\alpha(\rho_{max})}
\frac{\rho}{(1-\rho/\rho_{max})^2} \, . 
\end{equation}
Note that, without further assumptions, 
$\chi(\rho)$ is determined in the single-lane model only
together with the overtaking probability 
$p(\rho)$. 
In our multi-lane model, we suggest the following decomposition of the
above expression:
\begin{equation}
  \chi_i(\rho_i) = 1 + \frac{V_{0i} T_i^2}{\tau_i\alpha_i(\rho_i^{\rm max})}
 \frac{\rho_i}{(1-\rho_i/\rho_i^{\rm max})^2} \, ,
\end{equation}
\begin{equation}
  p_i^a(\rho_i) = \frac{\exp{\left(-p_{0i}^a\rho_i/\rho_i^{\rm max}\right)}}
 {\chi_i(\rho_i)} \, . 
\end{equation}
Here, variables without an index for the specific vehicle type $a$
represent weighted averages of the variables belonging to 
the different vehicle types in lane $i$, 
for example, 
\begin{equation}
 T_{i} = \sum_{a=1}^A \frac{\rho_i^a}{\rho_i} T_{i}^a \, .
\end{equation}

\section{Calibration of the Multi-Lane Model and 
Simulation Results} \label{sec-calibrate}

Next, we present the results of calibration and simulation for a
special case of the general model discussed above. We consider the
two-lane variant of the model for a single vehicle type, where
$i=1$ represents the right (``slow'') lane and $i=2$ the left (``fast'')
lane. The calibration was done on the basis of empirical data for the 
Dutch two-lane highway A9.

\begin{figure}
  \begin{center}
    \includegraphics[width=0.5\textwidth]{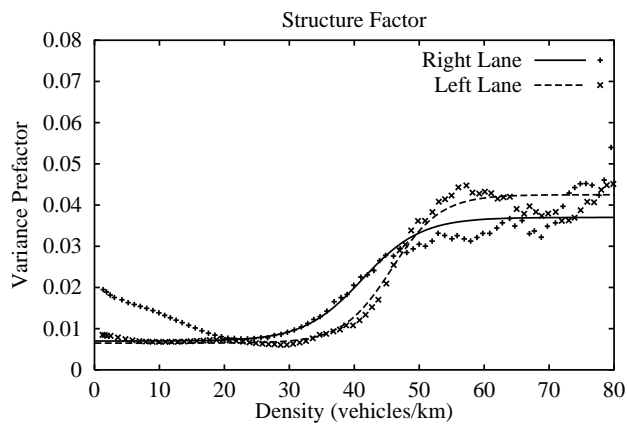}
  \end{center}
  \caption[]{ \label{fig-fitvar} Fit of the density-depenent 
variance prefactor or ``structure factor''
$\alpha_i(\rho_i)$ specified in Eq.~(\ref{eq-step}) (---) to
the empirical data of the relative velocity variance
$\theta_i/V_i^2$ (+, $\times$). The corresponding parameter values are
listed in Table~\ref{tab-fit}. Note that the deviation of the
fit curve for the right lane from the empirical data at small densities 
is probably a consequence of assuming one vehicle type only instead of
heterogeneous traffic. However, this deviation is of minor importance 
for the dynamic properties of the model, since it is limited to the
stable density regime of free traffic flow.} 
\end{figure}

\begin{figure}
  \begin{center}
    \includegraphics[width=0.5\textwidth]{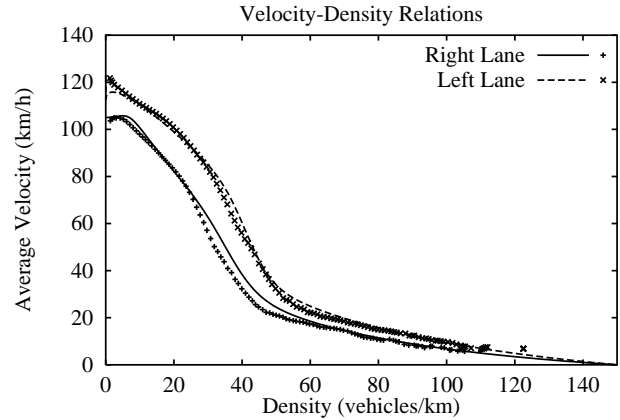}
  \end{center}
  \caption[]{ \label{fig-fitvel} Equilibrium 
velocity-density relations of the multi-lane model (---) for the
parameter values specified in Table~\ref{tab-fit} in comparison with
empirical data from the Dutch highway A9 (+, $\times$). The symbols
represent the (vertical) averages of one-minute data that were
evaluated for 14 successive days.} 
\end{figure}

The variance prefactor or ``structure factor''
$\alpha_i(\rho_i) = \theta_i / {V_i}^2$ can be estimated on the
basis of direct observation. The empirical data show a ``step-like''
but smooth dependence of this prefactor on density, with an
increase at about 40 vehicles per kilometer. 
It can be well fitted by the following
function (see Fig.~\ref{fig-fitvar}):
\begin{equation}
  \alpha_i(\rho_i) = \alpha_{0i} + \Delta \alpha_i \left[ 1 +
    \exp{\left( -\frac{\rho_i-\rho_{{\rm c}i}}{\delta\rho_i}\right)} 
\right]^{-1} \, . \label{eq-step}
\end{equation}
This step-like form plays an important role, as it determines the specific
shape of the equilibrium 
velocity-density relation $V_i^{\rm e}(\rho_i)$ (see Fig.~\ref{fig-fitvel})
and the fundamental diagram $Q_i^{\rm e}(\rho_i) 
= \rho_i V_i^{\rm e}(\rho_i)$. Other
parameters that influence the fundamental diagram are 
the safe time headway $T_i$, the desired velocity $V_{0i}$, and the
maximum density $\rho_i^{\rm max}$. The desired velocity fits the
maximum velocity in free traffic, whereas the safe time headway and
the maximum density determine the slope of the fundamental diagram 
at high densities and its intersection point with the density axis.
The acceleration relaxation
times $\tau_i$ and the anticipation factors $\gamma_i$ do not influence
the fundamental diagram. Instead, they allow to fit the stability behavior 
and dynamics of traffic flow~\cite{THH-98}. 

The parameters that affect lane-changing processes are the
coefficients $g_{i,j}$, $\beta_1$, and $\beta_2$ 
for the spontaneous lane-changing
rates, and the coefficients $p_{0i}$ for the overtaking probabilities.
The main sources for estimating these parameters are empirical data on lane
occupancies and lane-changing rates 
as a function of density. 
One can easily obtain the data on lane occupancies from the usual
measurements made by induction loops. In contrast, the direct
measurement of lane-changing events is much more difficult, which results
in a lack of reliable data in the literature 
\cite{Br-McD-96}. Luckily, this kind of data is only necessary for the
estimation of the order of magnitude of the model coefficients, while the
ratios of the coefficients for different lanes, which essentially define the
multi-lane dynamics, can be well estimated by the available lane
occupancy data.

\begin{figure}
  \begin{center}
    \includegraphics[width=0.5\textwidth]{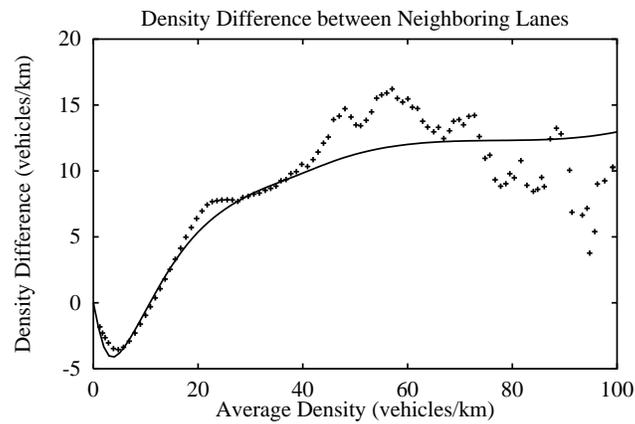}
  \end{center}
  \caption[]{ \label{fig-dendiff} Theoretical (---) and empirical (+)
difference of the
densities in the left and the right lane as a function of the lane-averaged
density. }
\end{figure}

\begin{figure}
  \begin{center}
    \includegraphics[width=0.5\textwidth]{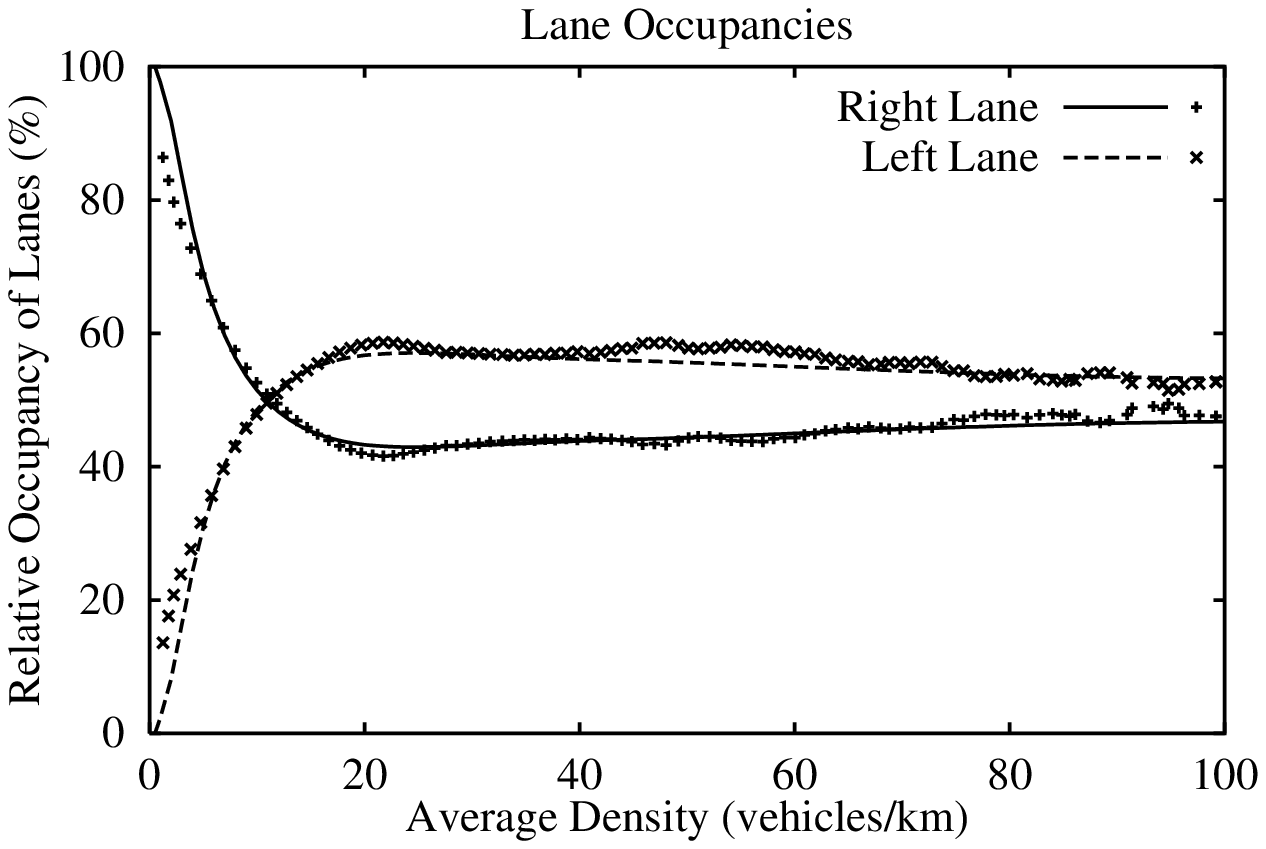}
  \end{center}
  \caption[]{ \label{fig-fitoccup} Fit of the lane occupancy, i.e. of
the percentage of the vehicle density in a lane compared to the 
total density in all lanes. Symbols correspond to empirical data,
lines to the results of our multi-lane model. The preference for the
right lane at small densities comes from the European
traffic regulations. }
\end{figure}

\begin{figure}
  \begin{center}
    \includegraphics[width=0.5\textwidth]{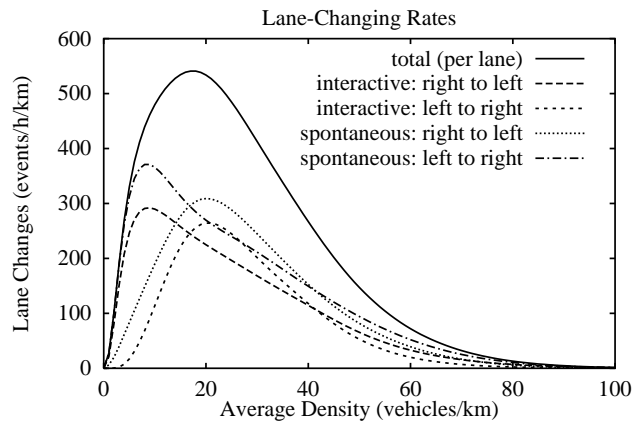}
  \end{center}
  \caption[]{ \label{fig-lcrate} Equilibrium lane-changing rates 
according to our multi-lane model as a function of the average density. 
Note that, at small densities, we have more interactive lane changes from
the right to the left lane (``overtaking maneuvers'')
and more spontaneous lane changes from the
left to the right lane, as expected for Europe.}
\end{figure}

Figures~\ref{fig-dendiff} through \ref{fig-lcrate} 
show the corresponding fits.
Following Ref.~\cite{Sparmann}, we assume that
the maximum lane changing rate of about 500 to 550 events per hour, kilometer
and lane is achieved at
densities of about 20 to 25 vehicles per 
kilometer. The higher occupancy of the 
left lane at middle and high densities yields higher estimated values of
the coefficients $g_{i,j}$, $p_{0i}$ for the left lane relative to those
for the right lane. The primary use of the right
lane at small densities (see Fig.~\ref{fig-dendiff})
reflects the European traffic regulations. One can
take this into account 
by a ``European-rules'' correction prefactor of the spontaneous 
coefficients:
$g_{12} \rightarrow g_{12} g_{\rm Eu}(\rho_1), g_{21} \rightarrow g_{21}/
g_{\rm Eu}(\rho_1)$, where $0 < g_{\rm Eu}(\rho_0) < 1$ 
is a smooth step-like function similar to~(\ref{eq-step}).

The calibration results show that the spontaneous lane-changing terms influence
mainly the low-density regime, while the lane changes due to interactions,
which are negligible at small densities, determine the difference in lane
occupancy at high densities. For the typical fit of the lane occupancy curves
(see Fig.~\ref{fig-fitoccup}) it turned out that the overall contributions of 
spontaneous and interactive lane-changing terms to the total lane-changing
rate were approximately the same
(Fig.~\ref{fig-lcrate}). Note that, at low densities, lane changes correspond
mainly to interactive lane changes from the right to the left 
lane and to spontaneous lane changes from the left to the right
lane, which is plausible for European traffic.

The results of the parameter estimation are summarized in 
Table~\ref{tab-fit}.

\begin{figure}
  \begin{center}
    \begin{minipage}{160mm}
      \includegraphics[width=78mm]{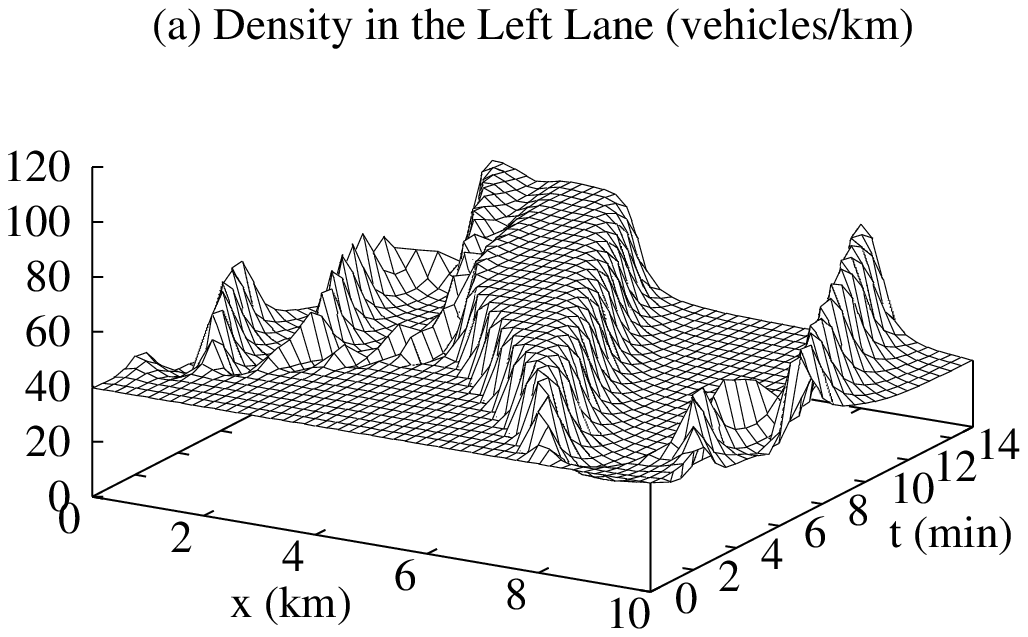}
      \includegraphics[width=78mm]{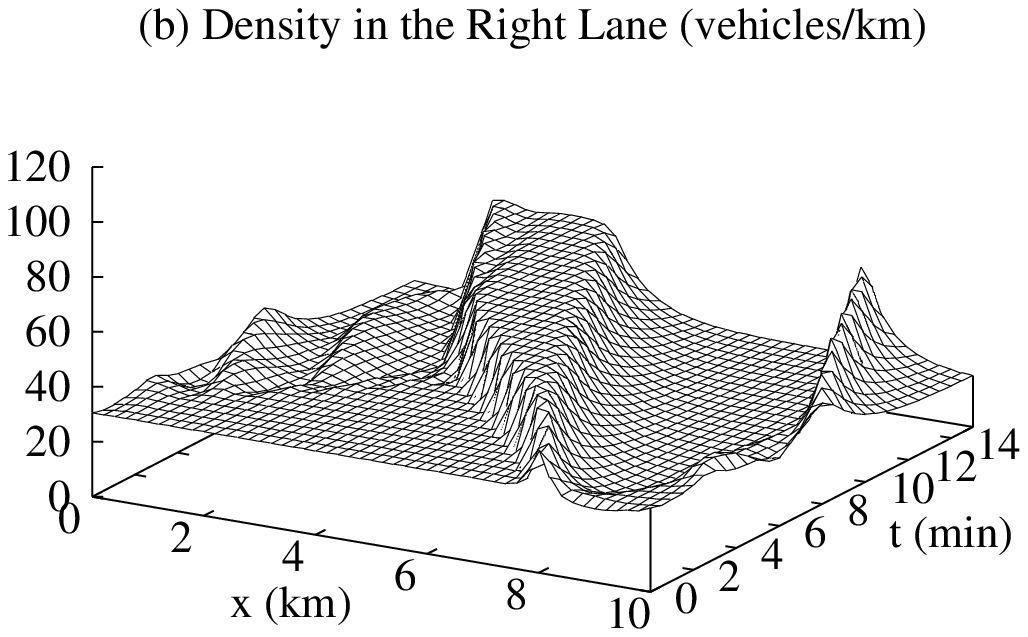}
      \includegraphics[width=78mm]{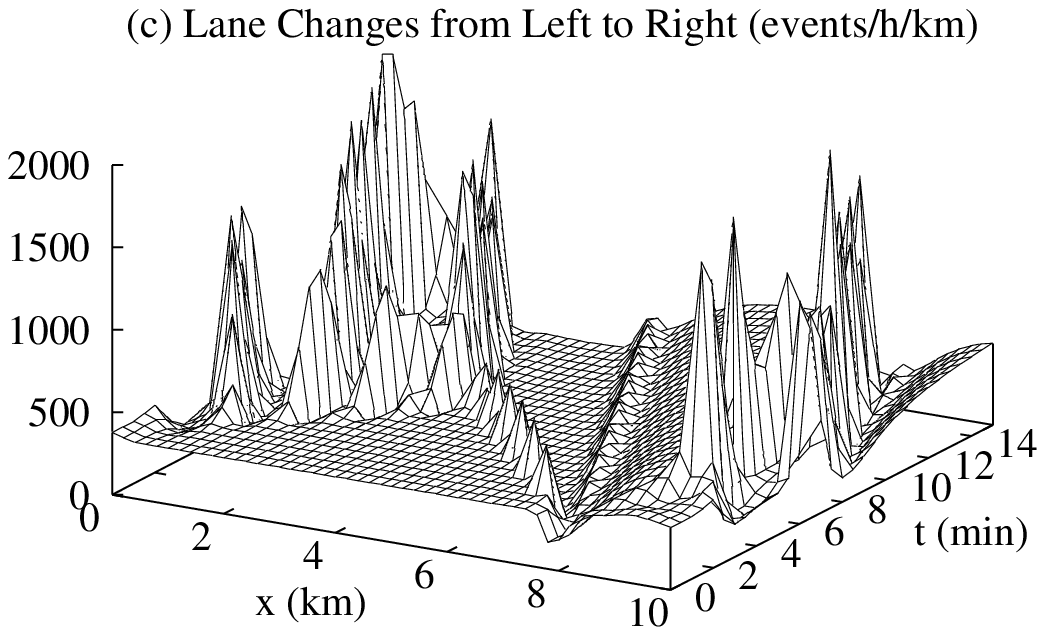}
      \includegraphics[width=78mm]{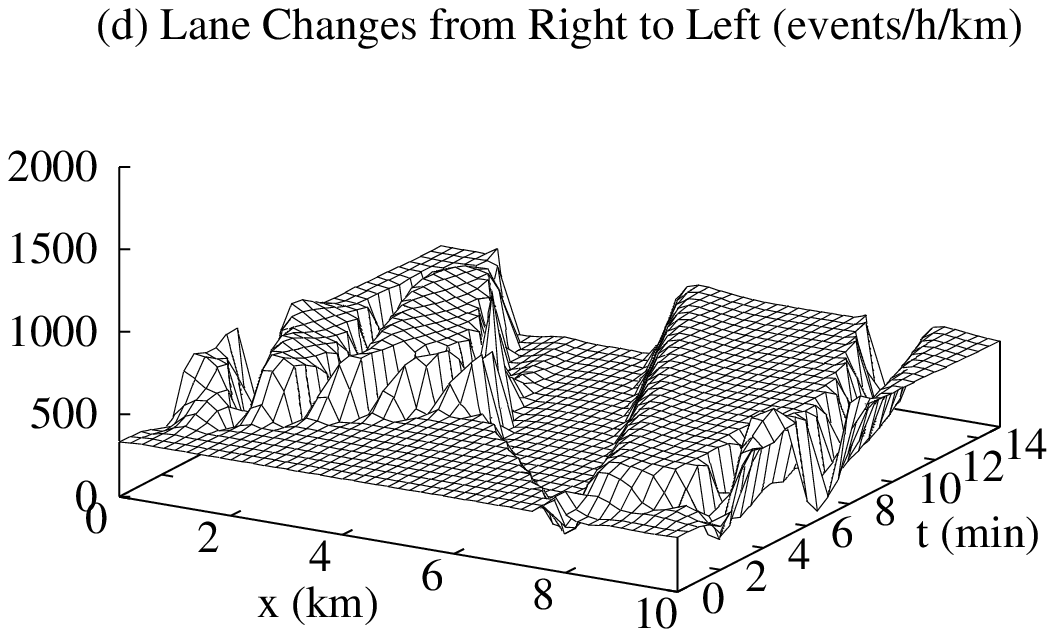}
    \end{minipage}
  \end{center}
  \caption[]{ \label{fig-sng} (a)+(b): Simulation of stop-and-go traffic 
  on a circular road in the
  regime of unstable traffic flow, arising from an initial
  density perturbation in the right lane, which eventually spreads to 
  the left lane. 
  (c)+(d): The lane-changing rates have temporary peaks
  at the locations, where the traffic situation in the neighboring
  lanes evolves differently. This tends to reduce the differences
  among lanes, so that similar spatio-temporal 
  traffic patterns form in both lanes. Consequently, 
  we have a synchronization of lanes at medium and high vehicle
  densities.}
\end{figure}

One important property of the above model is the ability to describe the
development of different congested traffic
states \cite{science,inhom}. Figure~\ref{fig-sng} shows the development of
stop-and-go traffic, which arises from a small density perturbation 
in the right lane. Due to lane changes, the perturbation spreads to
the other lane, and the traffic dynamics on the neighboring lanes
becomes synchronized \cite{KK-96,KK-97,Lee}. In particular, this holds for the 
propagation of large density clusters.
Nevertheless, the traffic flow in the left lane behaves more 
unstable in the range of moderate densities.
This fact is in agreement with observations and can be theoretically 
explained by the different velocity-density relations 
(decreasing more rapidly for the left lane,
see Fig.~\ref{fig-fitvel}).

Next, we present multi-lane simulations of the interesting case of 
a bottleneck, corresponding to an on-ramp, a lane closure, or an accident. 
Assume, for example, that the right lane 
ends at certain place $x_{\rm end}$. We expect that
the resulting traffic situation
will depend on the volume of the incoming flow.

In order to model the behavior of traffic close to a bottleneck, 
we must specify the mandatory lane changes. 
In the framework of the above multi-lane model, this can be done by
introduction of additional lane-changing terms describing 
a sufficient increase of lane changes to the left lane, whereas
lane changes to the right lane will be surpressed by setting the
corresponding coefficients close to the bottleneck to zero.
However, the
following difficulty arises: While the density on the right lane
decreases to zero at the bottleneck, the velocity (which depends on
the density and velocity on the neighboring lane) can stay large up to
the very end of the lane. This causes numerical
problems in keeping the density and flow positive everywere. To avoid
this, we apply the following calculation procedure to
the last section of the right lane in front of the
bottleneck, where we have assumed that this section 
is of length $L_0 = 500$\,m throughout this paper.

Close to the bottleneck, all drivers in the right lane 
(producing a traffic flow of volume $\rho_1V_1$)
must merge into the
adjacent lane. This implies that the drivers in the right lane will adopt
their velocity $V_1(x,t)$ to the velocity $V_2(x,t)$ 
in the left lane. In addition, we will assume that the lane-changing
rate grows inversely proportional to the remaining distance $L(x)=(x_{\rm
  end}-x)$, in order to guarantee that all vehicles have changed
lane at the place $x_{\rm end}$ where the right lane ends.
Hence, for $x \in [x_{\rm end} - L_0,x_{\rm end}]$,
our model for the right lane reads:
\begin{equation}
  \frac{\partial \rho_1}{\partial t} + 
  \frac{\partial (\rho_1 V_1)}{\partial x} = - \frac{\rho_1 V_1}{L(x)} \, , 
\end{equation}
\begin{equation}
  V_1(x,t) = V_2(x,t) \, .
\end{equation}
For the left lane, we have:
\begin{equation}
 \frac{\partial \rho_2}{\partial t} + \frac{\partial (\rho_2V_2)}
 {\partial x} = \frac{\rho_1 V_1}{L(x)} \, .
\end{equation}
In order to describe a smooth transition from the ``normal'' to this
``adaptive'' behaviour in the merging zone, we evaluate the RHS of the
equation according to
\begin{equation}  
  {\rm RHS} = [1-k(x)] {\rm RHS}_{\rm norm} + k(x) {\rm RHS}_{\rm adapt} \, , 
\end{equation}
where $k(x)$ is a smooth step-like function similar to~(\ref{eq-step})
with $k(x_{\rm end}-L_0) \approx 0$ and $k(x_{\rm end}) \approx 1$.
 
\begin{figure}
  \begin{center}
    \begin{minipage}{160mm}
      \includegraphics[width=78mm]{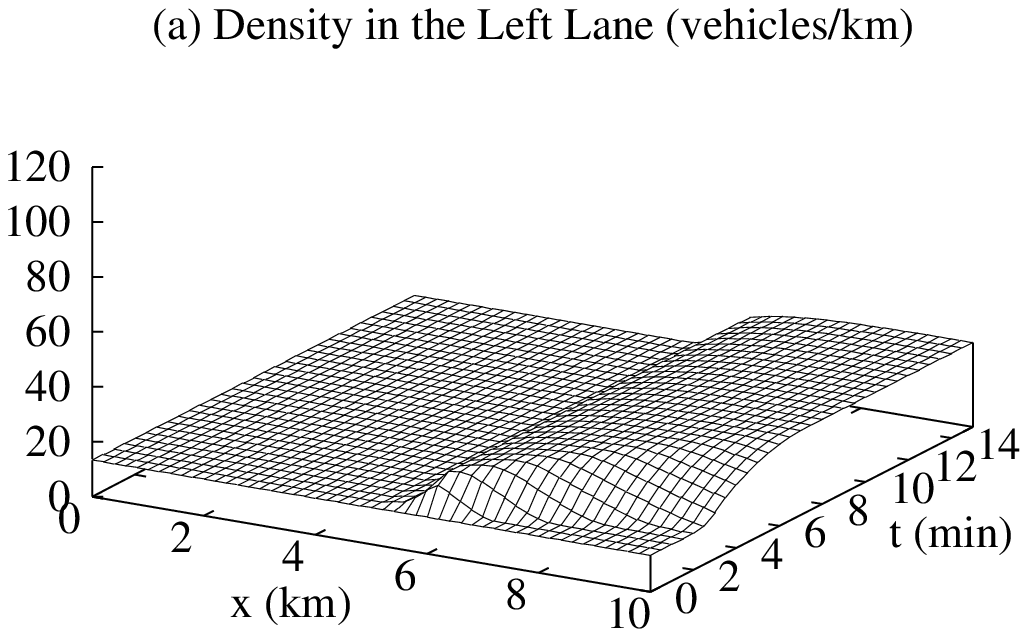}
      \includegraphics[width=78mm]{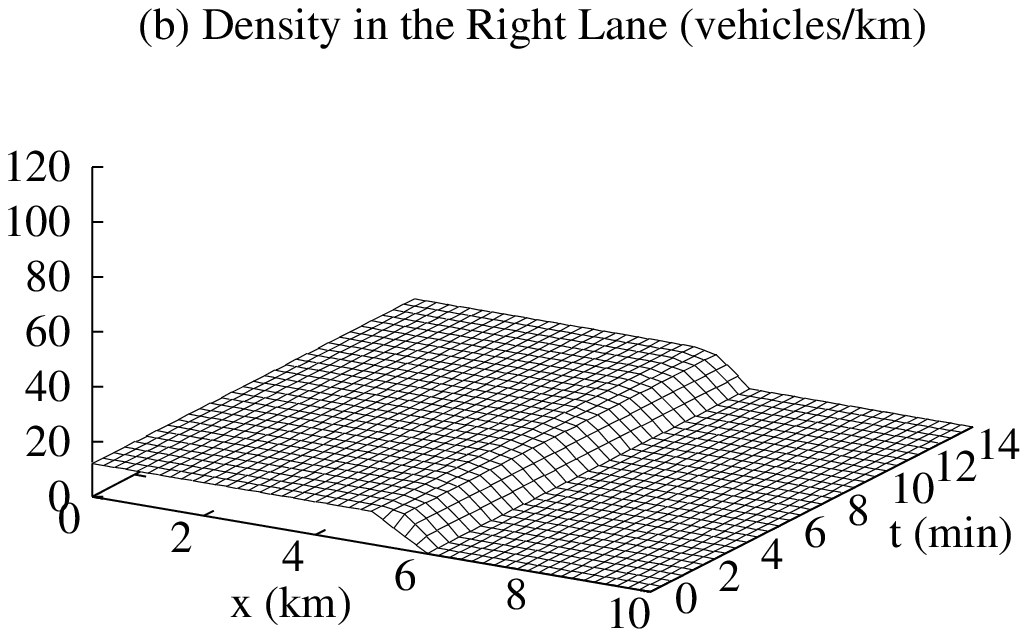}
    \end{minipage}
  \end{center}
  \caption{ \label{fig-lc10} Simulation of a closure of the right lane 
for $x \ge x_{\rm end} = 6$\,km at time $t=0$\,min. 
At a density of $\rho_{\rm init} = 12.8$ vehicles per
kilometer and lane or lower (here: $\rho_{\rm init} = 12.6$\,vehicles/km), 
the capacity of the left lane is large enough to
transport the vehicle flow from both lanes, resulting in a higher
vehicle density in the left lane downstream of the bottleneck,
whereas the right lane is empty behind the lane closure.}
\end{figure}

\begin{figure}
  \begin{center}
    \begin{minipage}{160mm}
      \includegraphics[width=78mm]{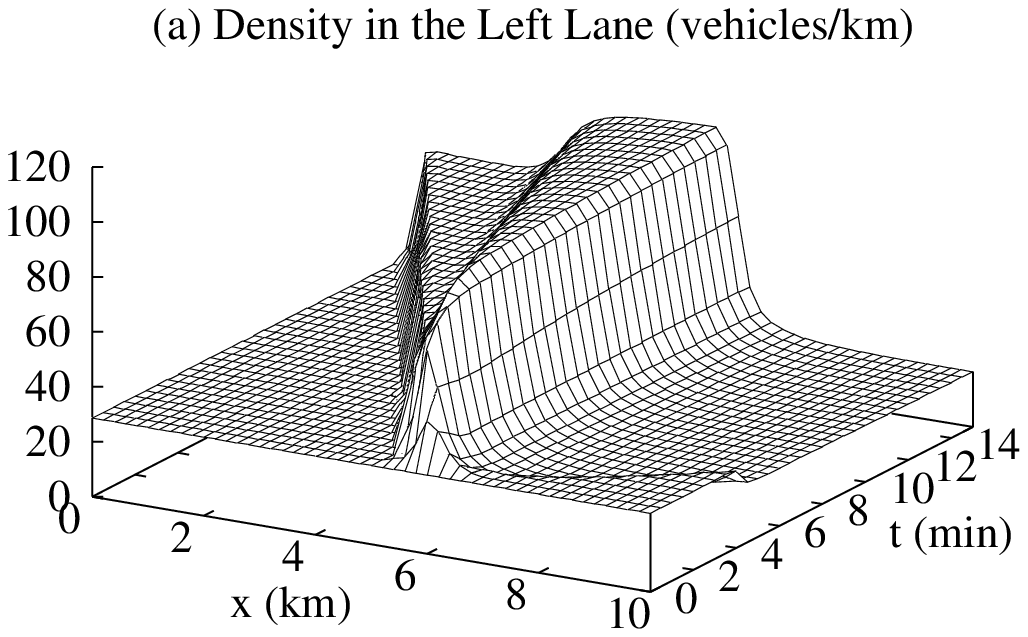}
      \includegraphics[width=78mm]{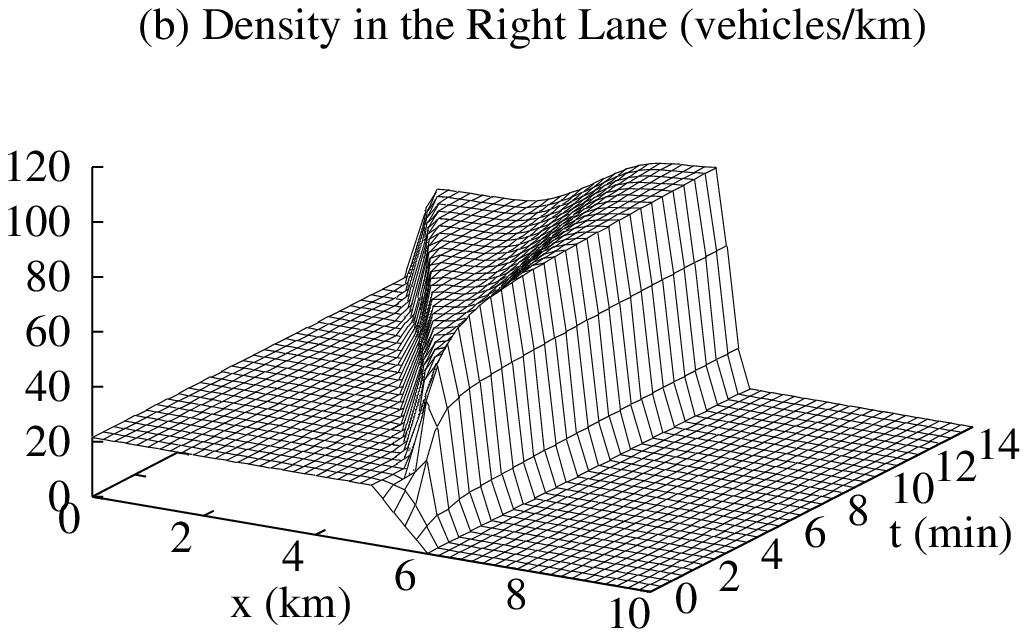}
      \includegraphics[width=74mm]{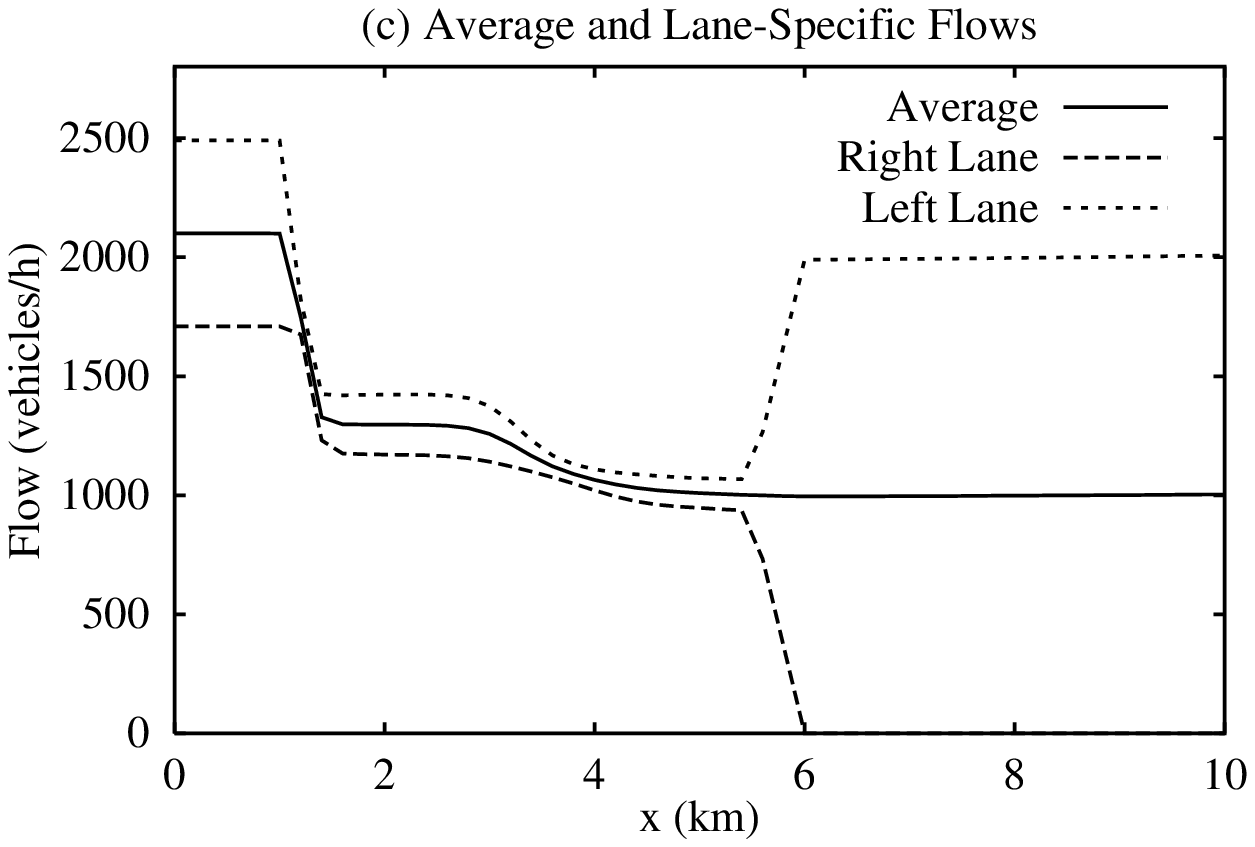}
\vspace*{3mm}\includegraphics[width=74mm]{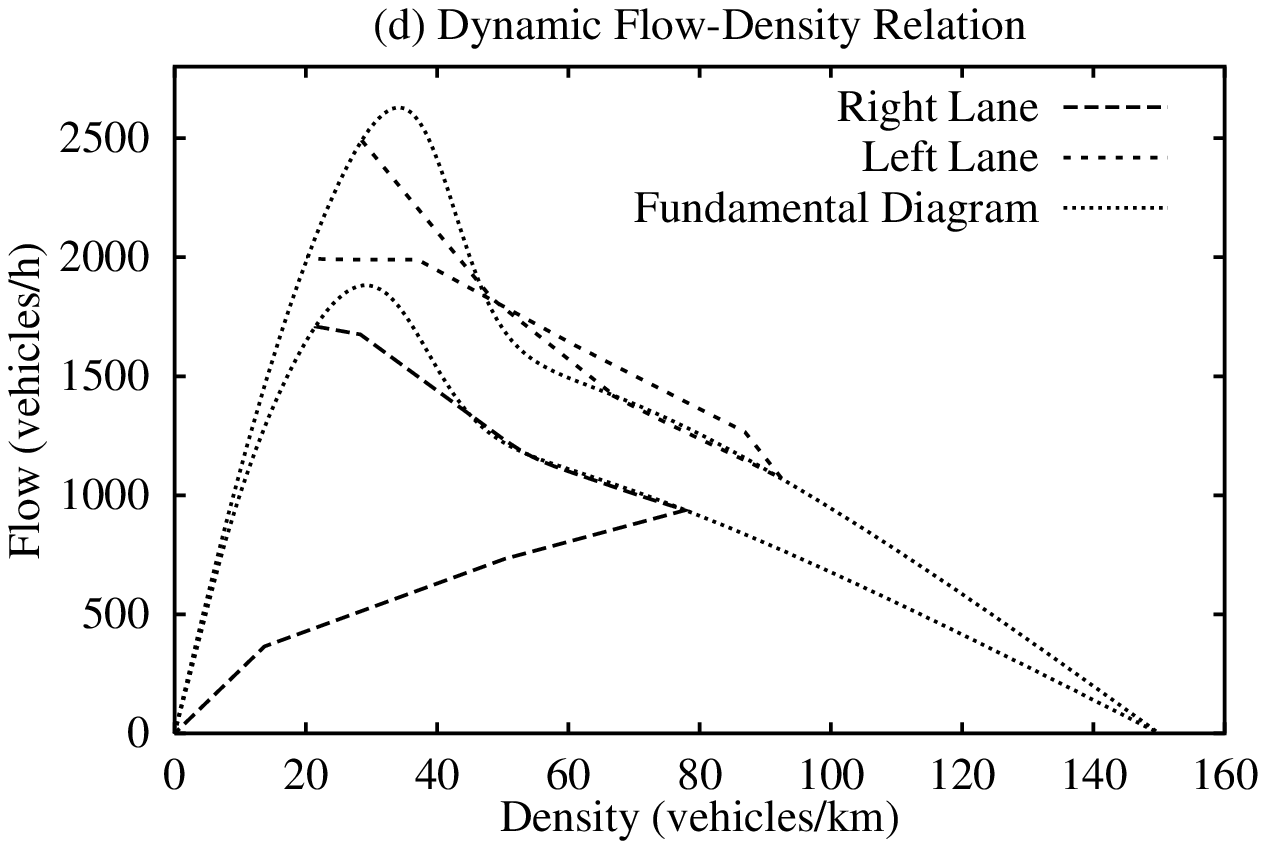}
    \end{minipage}
  \end{center}
  \caption[]{ \label{fig-lc25} (a)+(b): Closure of the right lane as in
    Figure~\ref{fig-lc10}, but for an average initial density of
$\rho_{\rm init} = 25$ vehicles per kilometer and lane. Behind the
bottleneck, a growing region of congested traffic forms immediately,
since the capacity of the left lane is exceeded by the traffic volume
in both lanes. (c) Note that the effective capacity of the left lane
(i.e. of the bottleneck) is considerably less than the maximum flow 
and the flow in the left lane upstream of the jam. This is, because the
outflow from congested traffic is a self-organized quantity
\cite{THH-98}, which is of the order of 2000 vehicles per hour, here.
The flow per lane in the jammed region is half of this characteristic
outflow. (d) The step-like structure of congested traffic
corresponding to regions of two different densities 
in (a) and (b) is related to
a deceleration in two steps (rough braking and fine braking), when
approaching a traffic jam from free traffic. This behavior has
been also observed in a microscopic traffic model \cite{HeTi98}.
According to an explanation by Ansgar Hennecke, it relates to the 
pronounced hump of the fundamental diagram in the density region between 20
and 50 vehicles per kilometer. 
Where the traffic flow is stable (at densities around
60 vehicles per kilometer and higher), the flow-density relation tends
to stay close to the fundamental diagram 
(i.e. the {\em equilibrium} flow-density
relation). In contrast, the dynamic flow-density relation is
a self-organized relation in the density regime of unstable traffic
flow (at densities between about 25 and 55 vehicles per kilometer), 
connecting the stable flow with the self-organized outflow from
traffic jams. Hence, we will usually have different slopes in the
resulting dynamic flow-density relation, corresponding to the
propagation of congested regimes of different densities with different
speeds. This behavior disappears for smoother fundamental diagrams,
in which the congested part decreases more or less linearly.}
\end{figure}

\begin{figure}
  \begin{center}
    \begin{minipage}{160mm}
      \includegraphics[width=78mm]{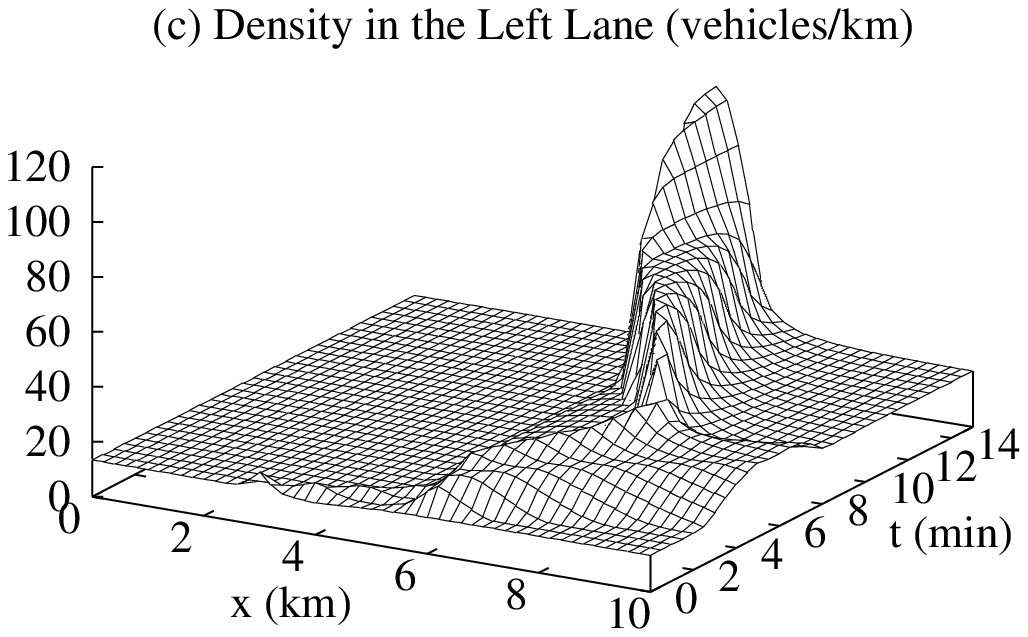}
      \includegraphics[width=78mm]{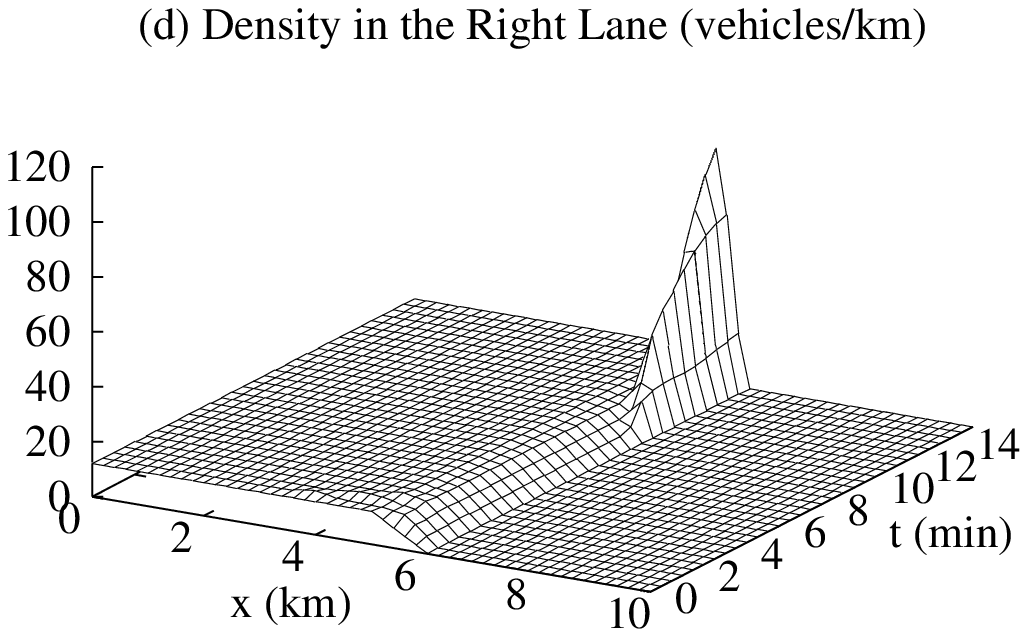}
    \end{minipage}
  \end{center}
  \caption[]{ \label{fig-lc15} Simulation of a closure of the right lane as
in Figure~\ref{fig-lc10}, with the same initial density of
$\rho_{\rm init} = 12.6$ vehicles per kilometer and lane, but with a
small perturbation of the traffic flow in the left lane. 
Altough the total flow of 2590 vehicles per hour in both lanes 
is below the maximum possible flow 
in the left lane of 2630 vehicles per hour, 
traffic flow eventually breaks down.
In other words: If all vehicles would use only the left lane,
there would be no traffic congestion upstream of $x_{\rm end}$! 
The breakdown of traffic is initialized by a
perturbation of traffic flow that eventually 
gives rise to a growing region of congested traffic \cite{HT-98}, 
from which the
self-organized outflow is only about 2000 vehicles per hour \cite{THH-98}.}
\end{figure}

The results of our simulations are presented in
Figures~\ref{fig-lc10} through \ref{fig-lc15}.
The traffic dynamics is essentially
characterized by the volume of approaching traffic and
the capacity of the bottleneck (which is given by the
outflow from traffic jams in the left lane).
At low upstream densities (see Fig.~\ref{fig-lc10}), the capacity of the left
lane is sufficient to transport the vehicle flow 
from both lanes. In contrast, we have an immediate formation of
congested traffic upstream of the bottleneck, if the total traffic volume
in both lanes exceeds the capacity of the left lane
(cf. Fig.~\ref{fig-lc25}). Surprisingly, for a certain range of 
moderate densities,
the resulting traffic situation turns out to depend on the initial
condition. While a perfectly homogenous flow will lead to
an increased but free traffic flow downstream of the bottleneck, as for
small traffic volumes, a small perturbation can trigger the
breakdown of traffic flow, although the left lane could carry
the total vehicle flow in both lanes (cf. Fig.~\ref{fig-lc15}). 
This can happen, when the
traffic flow downstream of the bottleneck is unstable. As long as the
perturbation is small, it moves downstream. However, when its amplitude
becomes larger, it eventually changes its propagation speed and
finally travels upstream, until it reaches 
the bottleneck. Then, traffic breaks down, and a steadily growing 
region of congested traffic develops upstream of the bottleneck,
whereas traffic downstream of the bottleneck flows freely
(Fig.~\ref{fig-lc15}). A similar phase transition from free to congested
traffic is known to occur close to on-ramps \cite{HT-98,KK-97}. 

\section{Comparison with the Effective Single-Lane Model}

The traffic situations discussed above can be also simulated
with an ``effective'' single-lane model that implicitly averages 
over the dynamics of all lanes. The corresponding model was
proposed in \cite{HT-98,THH-98} and basically corresponds to
our multi-lane model, applied to one lane only, so that the lane changing
terms drop out. For our simulations, we use the following ``effective''
model parameters: $V_0 = 110$\,km/h, $\rho_{\rm max} = 150$\,vehicles/km,
$\tau = 35$\,s, $T=1.6$\,s, $\gamma = 1.2$, $\alpha_0=0.007$,
$\Delta \alpha = 0.031$, $\rho_{\rm c}=0.28\rho^{\rm max}$, and
$\delta \rho = 0.025\rho^{\rm max}$. 

\begin{figure}
  \begin{center}
    \begin{minipage}{160mm}
      \includegraphics[width=78mm]{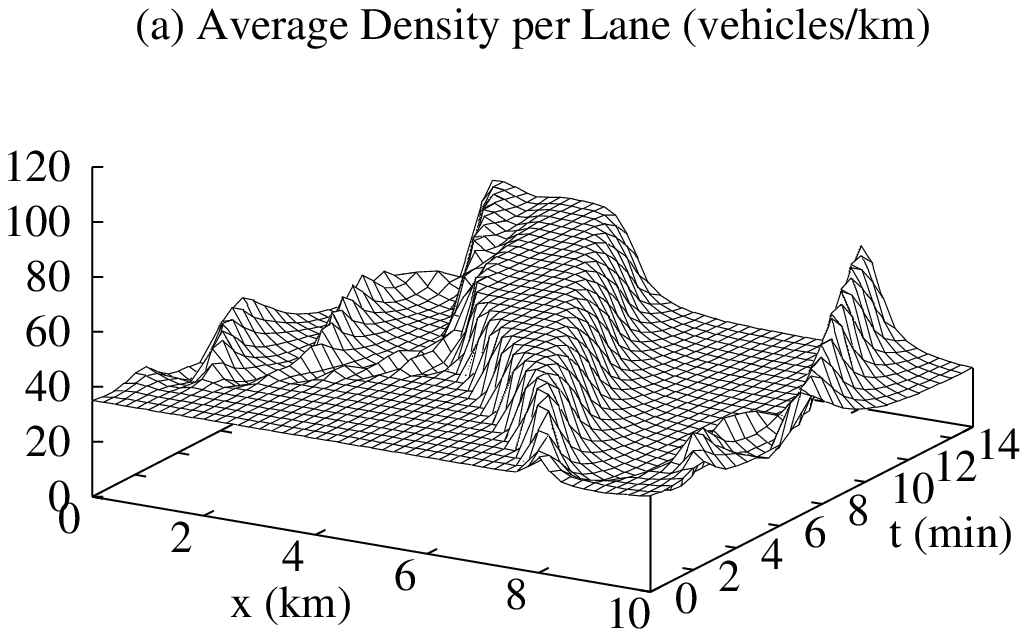}
      \includegraphics[width=78mm]{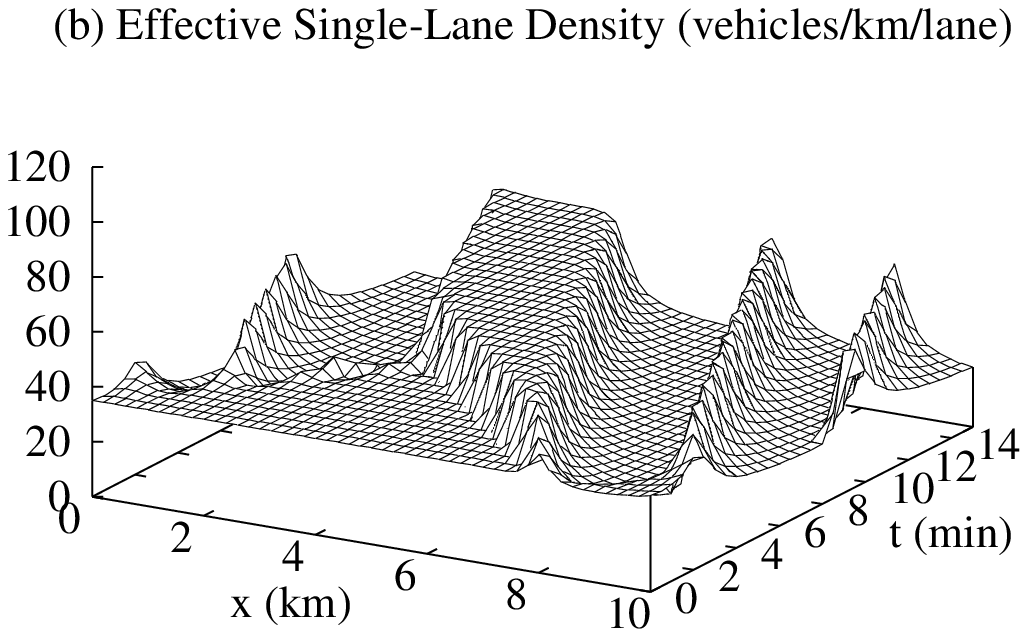}
    \end{minipage}
  \end{center}
  \caption[]{ \label{added} Comparison of (a) the average density 
according to the multi-lane model and (b) the density resulting from
an effective single-lane model for the formation of stop-and-go
traffic presented in Figure~\ref{fig-sng}.} 
\end{figure}

In Figure~\ref{added}, we
compare the average of the densities in the left and the right lane
according to the multi-lane model (see Fig.~\ref{fig-sng})
with the effective single-lane
model. It turns out that, despite of the sensitive dynamics in 
the unstable traffic regime, both models produce similar
spatio-temporal traffic patterns, but there are some differences
in detail. This shows that the effective single-lane model gives already a
reasonable representation of the traffic dynamics, although it produces small
deviations from the dynamics predicted by the multi-lane model.

Let us make a similar investigation for the example of
a bottleneck. In this case, we can treat the merging lanes in the
effective single-lane model by a reduction of the effective lane
number $I(x)$ from 2 to 1 within the merging section, i.e. for
$x \in [x_{\rm end} - L_0, x_{\rm end}]$. For example, we may use
the linear relation $I(x) = [1 + L(x)/L_0]$. The conservation of the
number of vehicles implies the virtual ramp flow 
$\nu = - (\rho V/I) \, \partial I/\partial x$, which gives the following
continuity equation for the effective vehicle density per 
available lane:
\begin{equation}
 \frac{\partial \rho}{\partial t}
 + \frac{\partial (\rho V)}{\partial x} = - \frac{\rho V}{I(x)} 
 \frac{\partial I}{\partial x} \, .
\end{equation}
 
\begin{figure}
  \begin{center}
    \begin{minipage}{160mm}
      \includegraphics[width=78mm]{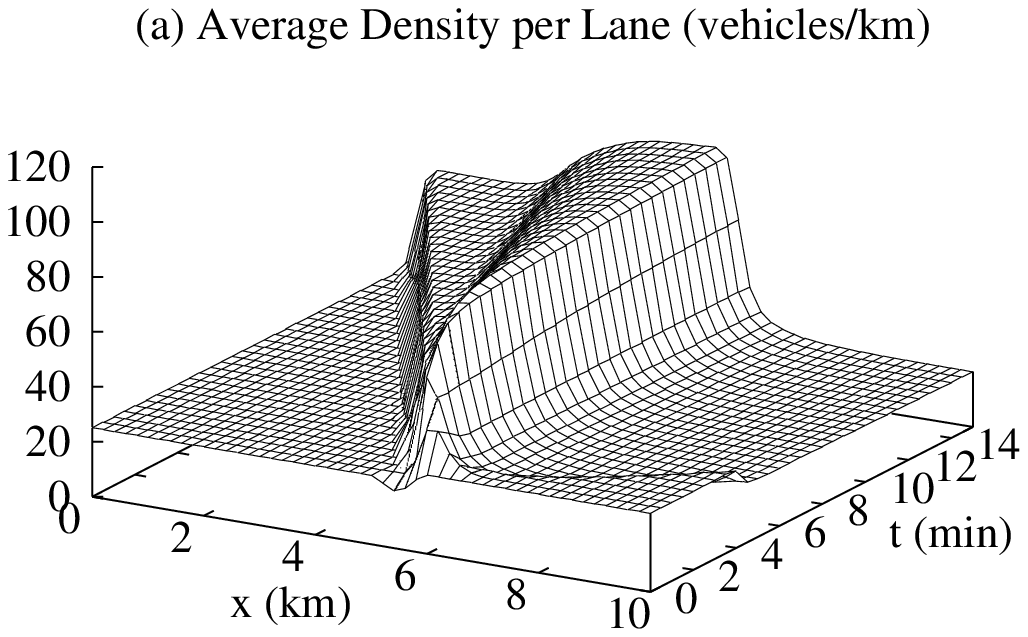}
      \includegraphics[width=78mm]{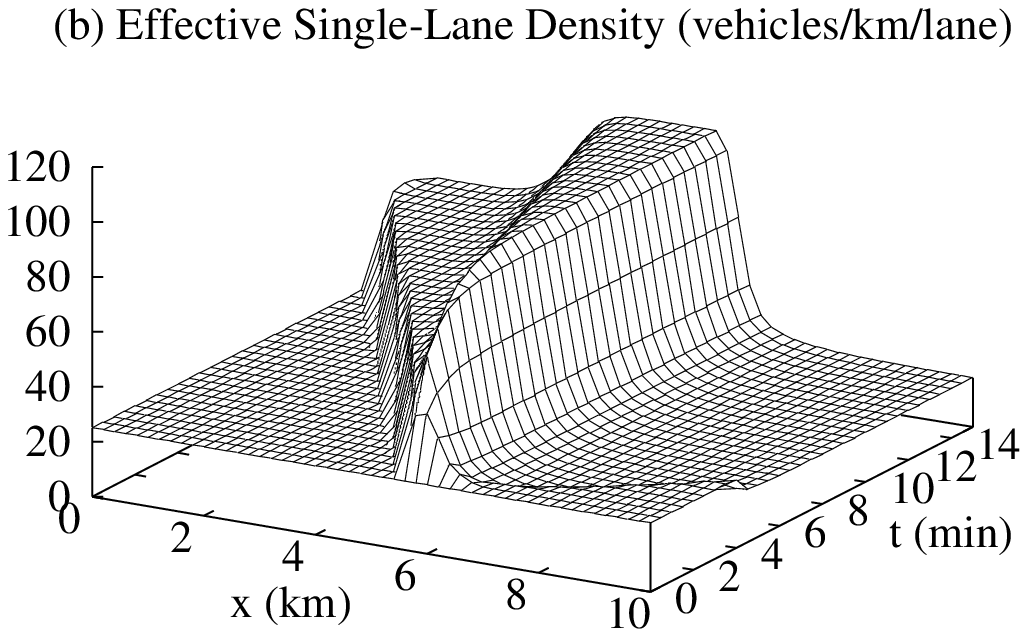}
    \end{minipage}
  \end{center}
  \caption[]{ \label{fig-single} Comparison of (a) the average density 
according to the multi-lane model and (b) the density resulting from
an effective single-lane model for the case of a lane closure
displayed in Figure~\ref{fig-lc25}.} 
\end{figure}

The result of the corresponding 
simulation is presented in Figure~\ref{fig-single} in
comparison with the plot of the average density per lane obtained with the
multi-lane model. The pictures
show a good correspondence between the multi-lane and the effective single-lane
models. However, there are slight differences in the form and propagation
velocity of the upstream front of the congested traffic region.
These originate from the fact that the average dynamics of two nonlinearly
behaving systems with different parameters
cannot simply be represented by one system of the same type
with suitably averaged parameters, as it can be done for linear systems.

\section{Summary and Conclusions}

We have proposed a gas-kinetic traffic model for heterogeneous multi-lane
traffic and systematically derived the corresponding
macroscopic traffic model. Thus,
effects of different vehicle types and lane changes are explicitly
taken into account. Whereas previous multi-lane models have usually assumed
spontaneous lane changes only, we managed to calculate the lane-changing rates
due to vehicle interactions and found that these are of the same order
of magnitude. Note that both, spontaneous and interactive lane
changes are necessary to describe the empirically observed 
density-dependence of the total lane changing rates, the lane occupancies,
and the density difference among lanes correctly.

Moreover, the multi-lane traffic
model formulated above treats vehicular space requirements and
high vehicle densities in the right way, and we have even discussed
possible effects of velocity correlations of interacting cars, which
basically reduce the interaction rates. The corresponding computer
simulations are robust also in the unstable traffic regime, so that
we did not need to eliminate the dynamic velocity equation, as
was done in a previous study. 

We have successfully calibrated our model to empirical traffic data.
The resulting model is in good agreement with the observed variance-density
relations, the velocity-density relations, and the occupancies of the
different lanes, as well as with the density-dependence of the
lane-changing rates and the density difference among lanes. We
were able to show the synchronization effect among lanes due to 
lane changes and could describe the traffic dynamics at bottlenecks.
A comparison of the average dynamics in the different lanes with
corresponding simulation results of an effective single-lane model
showed a qualitative, but not fully quantitative agreement.

Our present investigations focus on the empirical evaluation of 
velocity correlations between interacting vehicles and on the
calibration of the model to a mixture of vehicle types like cars and
trucks, both of which are difficult tasks. 
We expect that this will allow us to describe the effects of
heterogeneous traffic which were found in microscopic models
\cite{HH-98,zell} and other approaches \cite{scattered}.

\subsection*{Acknowledgments}
The authors want to thank for financial support by the BMBF (research
project SANDY, grant No.~13N7092) and by the DFG (Heisenberg scholarship
He 2789/1-1). They are also grateful to Henk Taale and the Dutch {\it
  Ministry of Transport, Public Works and Water Management} for supplying
the freeway data.

\begin{appendix}
\section*{}
In order to evaluate the macroscopic equations, one needs to calculate the
first two moments of the kinetic equation in velocity space. This
procedure was described in details in~\cite{Buch,Physica,drygran}. 
The new contribution of this
paper consists in evaluating the general Boltzmann
factors~(\ref{eq-boltz1}) through (\ref{eq-boltz3})
including vehicular space requirements and possible velocity correlations
of successive vehicles. The Boltzmann factors are defined by the
following integrals of interaction rates:
\begin{eqnarray}
A_i^{ab} &=& \int dv \; {\mathcal I}_i^{ab}(x,v,t) \, , \\
B_i^{ab} &=& \int dv \; v {\mathcal I}_i^{ab}(x,v,t)  
 - \int dv \; v {\mathcal J}_i^{ab}(x,v,t) \, , \\
C_i^{ab} &=& \int dv \; v {\mathcal I}_i^{ab}(x,v,t) \, .
\end{eqnarray}

Here, we present the evaluation of the integrals of 
${\mathcal I}_i^{ab}$ only, but the
integration of ${\mathcal J}_i^{ab}$ is analogous. 
For brevity, we will omit the indices of lane and
vehicle types in the following. 
This means that the pair distribution function
$f(x,v,x+s,w,t)$ actually denotes the pair distribution function
$f_i^{ab}(x,v,x+s_i^a,w,t)$ for particular types of vehicles
$a$ at point $x$ and $b$ at $(x+s_i^a)$ 
in a particular lane $i$. Consequently,
$V$, $\theta$ stand for $V_i^a(x)$, $\theta_i^a(x)$, and  $V'$, $\theta'$
stand for $V_i^b(x+s_i^a)$, $\theta_i^b(x+s_i^a)$. Also, we drop the multipliers
$\chi(\rho)\rho\rho'$. According to the definition of interaction
rates~(\ref{eq_intrate1}), one must evaluate the integrals
\begin{equation}
  {\mathcal B}_k = \int dv \int\limits_{v>w} dw \; v^k (v-w)
 f(x,v,x+s,w,t)\label{eq-integral}
\end{equation}
for $k\in\{0,1\}$.

From the mathematical perspective, the task is to integrate a bivariate
Gaussian distribution, multiplied by polynomials in $v$ and $w$, over the
half-plane $v>w$. This can be done by linear transformations in
the $(v,w)$-plane 
in three steps: 
\begin{itemize}
\item Transform the bivariate distribution to
the canonical rotation-symmetric form, 
\item rotate the plane to make
the boundary of the integration area parallel to one of the coordinate
axes, and
\item separate variables. 
\end{itemize}
Then,
the integration over one of the axes becomes trivial, and the integration
over the other axis gives combinations of terms which contain Gaussian and
error functions.

The bivariate Gaussian distribution $f$ is defined by~(\ref{eq-bivar1})
and (\ref{eq-bivar2}).
One can transform the quadratic function~(\ref{eq-bivar2}) to the simplest
symmetric form $v_1^2+w_1^2$ by linear transformation:
\begin{equation}
\left( \begin{array}{c} v \\ w \end{array} \right) =
\left( \begin{array}{c} V \\ V' \end{array} \right) + \mbox{\boldmath$C$}_1
\left( \begin{array}{c} v_1 \\ w_1 \end{array} \right) \, , \quad
\mbox{\boldmath$C$}_1 = \left( \begin{array}{ccc} 
\frac{1}{\sqrt{\lambda_+}}\cos\varphi & , &
-\frac{1}{\sqrt{\lambda_-}}\sin\varphi \\ 
\frac{1}{\sqrt{\lambda_+}}\sin\varphi & , &
\frac{1}{\sqrt{\lambda_-}}\cos\varphi
\end{array} \right) \, .
\end{equation}
Here, $\lambda_{\pm}$ are (positive) eigenvalues of the quadratic
form~(\ref{eq-bivar2}):
\begin{equation}
  \lambda_{\pm} = \frac{1}{2\theta\theta'(1-k^2)}
  \left( \theta+\theta' \pm \sqrt{\theta^2 - 2(1-2k^2)\theta\theta' +
\theta'^2} \right),
\end{equation}
and the angle $\varphi$ is defined by
\begin{equation}
  \tan\varphi = \frac{1}{2\sqrt{\theta\theta'}k}
  \left( \theta'-\theta - \sqrt{\theta^2 - 2(1-2k^2)\theta\theta' +
\theta'^2} \right).
\end{equation}
After this transformation, we obtain 
\begin{equation}
f(x,v,x+s,w,t)\;dv \, dw 
= \frac{1}{2\pi}e^{-\frac{1}{2}(v_1^2+w_1^2)}\;dv_1\, dw_1
\end{equation}
for the pair distribution function,
and the boundary of the integration area becomes
\begin{equation}
v-w = \frac{\cos\varphi-\sin\varphi}{\sqrt{\lambda_+}}v_1
- \frac{\cos\varphi+\sin\varphi}{\sqrt{\lambda_-}}w_1 + V-V' = 0 \,.
\end{equation}
Next, we apply additional rotation, which does not change the symmetric form 
of the distribution, and make the boundary of integration area parallel to
one of the axes, say $y$. This is done by
\begin{equation}
\left( \begin{array}{c} v_1 \\ w_1 \end{array} \right) 
= \mbox{\boldmath$C$}_2
\left( \begin{array}{c} x \\ y \end{array} \right) \, , \quad
\mbox{\boldmath$C$}_2 = \left( \begin{array}{ccc} 
\cos\psi & , & -\sin\psi \\ 
\sin\psi & , & \cos\psi
\end{array} \right),
\end{equation}
where
\begin{equation}
\cos\psi = \frac{\sin\varphi-\cos\varphi}{\sqrt{S\lambda_+}}, \quad
\sin\psi = \frac{\cos\varphi+\sin\varphi}{\sqrt{S\lambda_-}},
\end{equation}
\begin{equation}
S = \frac{(\sin\varphi-\cos\varphi)^2}{\lambda_+} + 
\frac{(\cos\varphi+\sin\varphi)^2}{\lambda_-} =
\theta - 2 k\sqrt{\theta\theta'} + \theta'.
\end{equation}
Finally, we consider the composition of two linear transformations: 
$\mbox{\boldmath$C$}=\mbox{\boldmath$C$}_2 \circ \mbox{\boldmath$C$}_1 =
( c_{ij} )$. 
This transformation brings the
integral~(\ref{eq-integral}) into the form with separating variables:
\begin{equation}
  {\mathcal B}_k = \int\limits_{x<\delta V} dx \;
\frac{e^{-\frac{x^2}{2}}}{\sqrt{2\pi}} 
\int\limits_{-\infty}^{+\infty} dy \;
\frac{e^{-\frac{y^2}{2}}}{\sqrt{2\pi}}  
\sqrt{S}(\delta V - x)(V+c_{11}x+c_{12}y)^k.
\end{equation}
The integration over $y$ becomes trivial, now, as it corresponds to 
evaluating the moments of a normal distribution. The evaluation for 
$k\in\{0,1\}$
involves only the first two moments, which equal to $1$ and $0$, hence 
the integration over $y$ in those cases results just 
in the elimination of the integral over
$y$ and $y$-containing terms from the expression above.

The coefficient $c_{11}$ reads
\begin{equation}
c_{11} = \frac{\cos\varphi\cos\psi}{\sqrt{\lambda_+}}
- \frac{\sin\varphi\sin\psi}{\sqrt{\lambda_-}} =
\frac{k\sqrt{\theta\theta'}-\theta}{\sqrt{S}}.
\end{equation}
Hence, for $k\in\{0,1\}$ we obtain
\begin{equation}
  {\mathcal B}_k = \int\limits_{x<\delta V} dx \;
\frac{e^{-\frac{x^2}{2}}}{\sqrt{2\pi}} 
S(\delta V - x)\left(\frac{V}{\sqrt{S}}+\frac{k\sqrt{\theta\theta'}-\theta}{S}x\right)^k.
\end{equation}

Note that the integration of ${\mathcal J}$, which is neccesary to obtain 
the Boltzmann factor $B$, leads to the same expression with the last
factor under the integral being replaced by
$\left(\frac{V'}{\sqrt{S}}+\frac{\theta'-k\sqrt{\theta\theta'}}{S}x\right)^k$.

The remaining task is to evaluate the ``incomplete moments'' of the normal
distribution, which can be expressed through the normal distributions ${N}(x)$ and
error functions ${E}(x)$ (see notations~(\ref{eq-norm}) and (\ref{eq-error})).
Applying the formulas
\begin{eqnarray}
\int\limits_{x<a} dx \; {N}(x) &=& {E}(a),\\
\int\limits_{x<a} dx \; x {N}(x) &=& -{N}(a),\\
\int\limits_{x<a} dx \; x^2 {N}(x) &=& -a{N}(a)+{E}(a)
\end{eqnarray}
to the integral above, one obtains the desired
expressions~(\ref{eq-boltz1}) through (\ref{eq-boltz3}) for 
the Boltzmann factors.
\end{appendix}

\begin{table}
\begin{tabular}{llll}
  Parameter & Notation & Right lane & Left lane\\ \hline
  Desired Velocity & $V_0$ & 105 km/h & 123 km/h\\
  Maximum Density & $\rho^{\rm max}$ & 150 vehicles/km & 150 vehicles/km\\
  Relaxation Time & $\tau$ & 35 s & 35 s\\
  Safe Time Headway & $T$ & 1.7 s & 1.2 s\\
  Anticipation Factor & $\gamma$ & 1.2 & 1.2\\
  Coefficients for Variance Approximation & $\alpha_0$ & 0.007 & 0.0065\\
  & $\Delta \alpha$ & 0.03 & 0.036\\
  & $\rho_{\rm c}$ & 0.275 $\rho^{\rm max}$ & 0.305 $\rho^{\rm max}$\\
  & $\delta\rho$ & 0.03 $\rho^{\rm max}$ & 0.025 $\rho^{\rm max}$\\
  Coefficient for Overtaking Probability & $p_0$ & 17.0 & 12.5\\
  Coefficient for Spontaneous Lane-Changing & $g_{i,3-i}$ & 75 & 28
\end{tabular}
\caption{The estimated parameter values for the two-lane, single
vehicle-class model, calibrated to traffic data from the Dutch motorway
A9. \label{tab-fit} }
\end{table}

%

\end{document}